\documentclass{article}

\usepackage{PRIMEarxiv}

\usepackage[utf8]{inputenc} 
\usepackage[T1]{fontenc}    
\usepackage{hyperref}       
\usepackage{url}            
\usepackage{booktabs}       
\usepackage{amsfonts}       
\usepackage{nicefrac}       
\usepackage{microtype}      
\usepackage{lipsum}
\usepackage{fancyhdr}       
\usepackage{graphicx}       
\graphicspath{{media/}}     
\usepackage{amsmath,amssymb}
\usepackage{enumitem}
\title{A note on the role of the initial error structure in the tropics on the seasonal-to-decadal forecasting skill in the extratropics
}
\author{
  St\'ephane Vannitsem \\
  Affiliation \\
  Royal Meteorological Institute of Belgium, Avenue Circulaire, 3, 1180 Brussels\\
  Belgium\\
  \texttt{Stephane.Vannitsem@meteo.be} \\
   \And
  Wansuo Duan\\
  Affiliation \\
  Key Laboratory of Earth System Numerical Modeling and Application, Institute of Atmospheric Physics, Beijing\\
  China\\
  \texttt{duanws@lasg.iap.ac.cn} \\
}

\begin{document}
\maketitle

\begin{abstract}
The predictability of a coupled system composed by a coupled reduced-order extratropical ocean-atmosphere model forced by a low-order 3-variable tropical recharge-discharge model, is explored with emphasis on the long term forecasting capabilities. Highly idealized ensemble forecasts are produced taking into account the uncertainties in the initial states of the system, with a specific attention to the structure of the initial errors in the tropical model. Three main types of experiments are explored with random perturbations along the three Lyapunov vectors of the tropical model, along the two dominant Lyapunov vectors, and along the first Lyapunov vector, only. When perturbations are introduced along all vectors, forecasting biases are developing even if in a perfect model framework. Theses biases are considerably reduced only when the perturbations are introduced along the dominant Lyapunov vector. This perturbation strategy allows furthermore for getting a reduced mean square error at long lead times of a few years, and to get reliable ensemble forecasts on the whole time range. These very counterintuitive findings further underline the importance of appropriately control the initial error structure in the tropics through data assimilation.
\end{abstract}

\keywords{Teleconnections, low-frequency variability, ENSO, predictability, chaos}

\section{Introduction}

When dealing with a weather or climate forecasting system, many different error sources may affect the quality of the forecasts. Two main sources of errors have been identified along the years, first the uncertainties in the initial state of the forecasting system and second the presence of uncertainties in the structure of the model \cite{Lorenz1982, Orrel2001, Vannitsem2002, Nicolis2003, Palmer2005, Prive2013, Duan2022}, whose dynamics are now quite well understood and taken into account in most state-of-the-art weather and climate forecasting systems \cite{Palmer2005, Buizza2010, Charron2010, Ollinaho2017, Jankov2017}. These uncertainties obviously affect directly the bulk of the forecasts, but additional sources of errors are present which also induce a loss of predictability. One important source are the ones coming from the boundaries of the domain considered, either at the interfaces with the other components of the climate system or from boundaries of limited area models \cite{Marsigli2014,Bouttier2016,Frogner2019}. The latter source was thoroughly investigated in a theoretical framework by \cite{Nicolis2007} who showed the complexity of the impact of such errors which considerably depend on the type of dynamics considered. She also indicated that the errors related to these boundaries would behave as a parametric model error in the numerical weather and climate prediction systems. A last source of uncertainty is related to the presence of external forcing in the forecasting system, which may also be affected by errors. An example of such a source of uncertainties is provided in \cite{Vannitsem2023} in which a coupled tropical-extratropical reduced-order model is considered. As for the boundary condition errors, these errors may be considered as parametric errors in numerical weather and climate prediction systems, but one should keep in mind that both boundary condition errors and external forcing errors may display a complex dynamics which may affect the predictability of the system in an unconventional way. Such a complication will be illustrated in the present paper in the context of an idealized system composed of two simplified coupled ocean-atmosphere models for the tropics and the extratropics.  

Teleconnections are known to be present between the tropical Pacific and the extratropics \cite{Philander1990, Alexanderetal2002, Fletcher2015, Stan2017, Schemmetal2018}. The presence of such influences suggest the possibility to gain in seasonal-to-decadal predictability in the extratropics as discussed for instance in \cite{Bronnimann2007, Senan2024}. In a recent study, we have explored in an idealized model setting the question of the gain of predictability one could expect from such teleconnections \cite{Vannitsem2023}. The model used is a tropical-extratropical coupled system, composed in the extratropics by a coupled ocean-atmosphere reduced-order model \cite{Vannitsem2015, DeCruzetal2016} and in the tropics by the low-order recharge-discharge oscillator model built along the years by Jin and colleagues \cite{Jin1996, Jin1997, Timmermannetal2003, Robertsetal2016, Guck2017}. In such a framework, the extratropical module is a nonautonomous system whose forcing is very slow as compared with the extratropical variability. It has been shown in particular that the presence of this tropical forcing indeed allows for an increase of predictability in the extratropics but the increase strongly depends on the amplitude of the errors present in the initial conditions in the tropical model, tempering the possibility to fully exploit the actual teleconnections. Furthermore, it was shown that even if a perfect model framework is adopted together with a perfect knowledge of the statistical properties of the initial errors encountered in both components of the system, the potential predictability (defined within the forecasting model only) is always larger than the actual skill of the forecasts (compared with the reference/observation system). This unexpected behavior was not addressed in this paper, but suggests that the link between potential predictability and actual skill is not only related to the presence of model errors or bad knowledge of the statistical properties of the initial errors as potential sources in the ongoing discussion on the signal-to-noise paradox \cite{Oreillyetal2018, CharltonPerez2019, Smithetal2020, Mayer2021, brocker2023}. The question of the source of discrepancy between the potential predictability and the actual skill in our tropical-extratropical system will be addressed in the present work, with the highly counterintuitive result that the long term forecasting reliability crucially depends on the nature of the initial errors in the tropical model. It will be shown in particular that when the initial error aligns along the most unstable direction, the agreement between the potential predictability and the actual skill is much better. 

In Section \ref{sec:model} the coupled tropical-extratropical system is described. Section \ref{sec:measures} introduces the different measures used to evaluate the predictability and skill of the forecasts, together with the experimental setup used. The results are the described in section \ref{sec:predictability}. Section \ref{sec:conclusions} is summarizing the key findings.

\section{The tropical--extratropical system}
\label{sec:model}

The system considered in the paper is a coupled tropical--extratropical model containing on one hand an extratropical reduced-order coupled ocean-atmosphere module and on the other hand, a low-order tropical module forcing the extratropical component. There is a one-way coupling from the tropics to the extratropics. The tropical and extratropical modules are presented below, together with the coupling introduced between them.

\subsection{The ENSO module}
\label{ssec:ENSO}

The ENSO model describes the dynamics of the temperature in the eastern and western tropical Pacific basins, together with the evolution of the thermocline depth \cite{Jin1996, Jin1997, AnJin2004, Timmermannetal2003, Robertsetal2016}. The model emulates the horizontal discharge-recharge mechanisms at play in the tropical Pacific through the heat exchanges between the tropical and subtropical waters, under surface wind stress and upwelling of subsurface cold water in the eastern part of the domain \cite{Jin1997, Robertsetal2016}. The equations are made non-dimensional  and reads (see \cite{Vannitsem2023}:      

\begin{subequations} \label{eq:ENSO}
\begin{align}
\frac{dx}{dt} & =  \rho \delta (x^2-a x) + s x (x+y+c-c \tanh(x+z)), \label{ENSO1} \\
\frac{dy}{dt} & =  -\rho \delta (a y+x^2), \label{ENSO2} \\  
\frac{dz}{dt} & =  \delta (k-z-\frac{x}{2}) \label{ENSO3}.
\end{align}  
\end{subequations}

with $x$ is the temperature difference between the eastern and western basins of the tropical Pacific, $y$ is the western basin's temperature anomaly with respect to a reference value, and $z$ is the western basin's thermocline depth anomaly. The dimensionless parameters are:

\begin{subequations} \label{eq:dimens}
\begin{align}
a & = \frac{ \alpha b L}{\epsilon h^{\*} \beta}, \qquad  \rho = \frac{\epsilon h^{\*} \beta}{r b L}, 
\qquad   \delta = \frac{r}{f_0}, \label{dimens-1} \\
c & = \frac{C}{S_0}, \qquad k = \frac{K}{S_0},  \qquad  s = \frac{\zeta h^{\*} \beta}{b L f_0}, \label{dimens-2} 
\end{align}  
\end{subequations}

with $f_0$ the Coriolis parameter used as time scale, and the other parameters as defined in \cite[Table~1]{Robertsetal2016}. For the applications presented in this paper, the set of parameter values used leads to a chaotic solution (Table \ref{tab:param}).

\begin{table}[ht!]
\centering
\caption{Dimensionless parameter values of the ENSO model}
\setlength\tabcolsep{6 pt} 
\begin{tabular}{l} 
\hline
\\
  $a=7.658609809$  \\
 $\rho =0.29016$       \\
 $\delta=0.0002803$     \\
$c=2.3952$   \\
 $k=0.4032$ \\
 $s=0.001069075$ \\
\hline
\end{tabular} \label{tab:param}
\end{table}

The ENSO model is the driving system in the coupled tropical-extratropical model. Equations~\eqref{eq:ENSO} can thus be integrated independently of the rest of the system, and this is done using a second-order Heun scheme with a time step $\Delta t = 0.1346$~hours $=0.05$ nondimensional time units. The variable $T_{ENSO}=x + y$ represents in the model \eqref{eq:ENSO} the sea surface temperatures in the eastern tropical Pacific that are commonly associated with the Ni\~no-3 index. A long chaotic solution was started from the initial state $(x = -2.8439, y = -0.62, z = 1.480)$ used in the sequel as the long reference trajectory of the tropical dynamics.

\subsection{The VDDG extratropical model and the coupling with the tropics} \label{ssec:VDDG}

The coupled ocean--atmosphere model used herein for the extratropics was developed by \cite{Vannitsemetal2015} and it is referred to hereafter as the VDDG model. Different versions of this model have been designed to explore the emergence of the low-frequency variabiliy (LFV) within the coupled ocean--atmosphere system \cite{DeCruzetal2016,Vannitsem2017}, and used as a multi-scale framework to analyze its instability properties \cite{VannitsemLucarini2016, DeCruzetal2018}, to construct stochastic parametrization \cite{Demaeyer2017, Demaeyer2018}, and to develop data assimilation schemes in coupled systems \cite{Pennyetal2019, Tondeuretal2020, Carrassietal2022}. The equations will not be repeated here as it has been considerably discussed in the publications just mentioned. The current configuration of the model is provided in \cite{Vannitsem2023}.

The extratropical model fields in both its atmosphere and its ocean are developed in Fourier series and truncated at a low order.  The number of modes herein is fixed at 10 for the atmosphere and 8 for the ocean, leading to 20 ordinary differential equations for the former and 16 for the latter. This model configuration is the original VDDG one; see also \cite{Vannitsem2017}. The parameter values used in the present work are the same as in \cite{Vannitsem2023}. In this configuration, there is no substantial low-frequency variability emerging in the extratropical component (See \cite{Vannitsem2022}), allowing for solely clarifying the impact of external low-frequency variability forcing.

As discussed in \cite{Schemmetal2018}, an important effect of the tropical forcing is to change the intensity of the zonal flow in both the North Pacific and the North Atlantic, with the impact of El Ni\~no and La Ni\~na differing from one region to the other. In order to mimic this dynamic effect in the extratropical VDDG model used here, a direct linear forcing of the model's first barotropic atmospheric mode as in \cite{Vannitsem2023}. This barotropic streamfunction mode represents the dominant term of the intensity of the zonal flow within the atmosphere. The impact of the forcing is therefore written as

\begin{equation}
\frac{d\psi_{a,1}}{dt} = f_1(\psi_{a,1}, \theta_{a,1}) + g \delta (x+y).
\end{equation} 

Here $f_1(\psi_{a,1}, \theta_{a,1})$ is the original right-hand side of the dynamical evolution of the first barotropic mode $\psi_{a,1}$; $\delta(x+y)$ represents the eastern tropical Pacific basin's temperature anomalies; and $g$ scales the intensity of the tropical forcing. The parameter $g$ represents the crucial forcing of  the midlatitude VDDG model described by the ENSO module described in Sec.~\ref{ssec:ENSO}. In our setting, given a positive $g$-value, a positive, warm anomaly will induce an increase of $\psi_{a,1}$,  and hence of the mean zonal flow, $U=- \psi_{a,1} \partial (\sqrt{2} \cos y) /\partial y$. This situation corresponds to the intensification of the zonal flow over the north Pacific during an El Ni\~no. If, to the contrary, $g$ is negative, this would correspond to an intensification during La Ni\~na that mimics the ENSO effect over the north Atlantic. In the current analysis, we are focusing on a few specific positive values of $g$ providing teleconnections between the tropics and the extratropics.

\begin{figure}
\centering
\includegraphics[width=140mm]{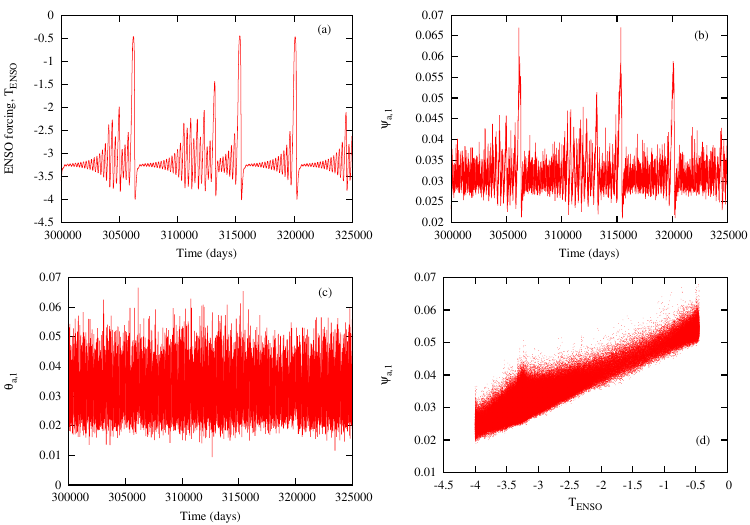}

\caption{Solutions of the coupled tropical-extratropical coupled ocean-atmosphere model for the coupling parameter $g=0.002$. (a) The tropical temperature forcing, $T_{ENSO}$; (b) the forced midlatitude variable, $\psi_{a,1}$; (c) the first baroclinic mode, $\theta_{a,1}$; and (d) the scatter plot of the variables $T_{ENSO}$ and $\psi_{a,1}$. This figure is reproduced from \cite{Vannitsem2023}.}

\label{solutions}
\end{figure}

Figure \ref{solutions} is reproduced from \cite{Vannitsem2023}, which displays a typical trajectory of the model for a coupling parameter of $g=0.002$. Panel (a) shows the forcing trajectory of sea surface
temperature in the east equatorial pacific with a succession of strong and less strong El-Niño, and la Niña events, occurring in an irregular manner. The evolution of
the first mode of the barotropic streamfunction, $\psi_{a,1}$, is shown in panel (b), indicating regular bursting associated with strong El-Niño events.
Panel (c) shows the same type of evolution for the first mode of the baroclinic streamfunction, $\theta_{a,1}$, strongly related to the evolution of $\psi_{a,1}$. In
this evolution, the bursting is not visible anymore, indicating a much lower influence of the tropical forcing than for  $\psi_{a,1}$. Finally in panel (d),
the relationship between the instantaneous values of $\psi_{a,1}$ and the temperature in the east part of the tropical ocean basin is displayed, indicating a strong teleconnection between the tropical Pacific and the streamflow of the midlatitude model. The correlation between the two observables is equal to 0.913, reflecting a
strong teleconnection between the two regions. If the tropical-extratropical coupling parameter $g$ is modified, one can either increase or
decrease the strength of these teleconnections. 




\section{Measures of predictability and experimental setup}
\label{sec:measures}

In the current experiments, initial errors of small amplitudes are introduced in the extratropical module, while errors in the tropical model are quite substantial. The main idea is to focus on the impact of errors in the external forcing. The impact of the errors in the external forcing will be present all along the error evolution as illustrated in \cite{Vannitsem2023}. We will focus first on the error at short time scales of a few days, and then to their impact at longer time scales typical of the evolution of the external forcing.

\subsection{Short time scales}
\label{shorttime}

Let us start by considering the evolution equations in an idealized form as
\begin{eqnarray}
\frac{d\bf{x}}{dt} & = & \bf{f}(\bf{x},\bf{y}, \lambda) \\
\frac{d\bf{y}}{dt} & = & \bf{s}(\bf{y},\zeta)
\end{eqnarray}
where the vector $\bf{x}$ of dimension $36$ contains the variables in the extratropical module and $\bf{y}$ of dimension $3$ the ones of the tropical module. The evolution of small amplitude errors is given by the linearized equations
\begin{eqnarray}
\frac{d\bf{\epsilon_x}}{dt} & = & \frac{\partial\bf{f}}{\partial \bf{x}}|_{\bf{x}, \bf{y}}   \bf{\epsilon_x} + \frac{\partial\bf{f}}{\partial \bf{y}}|_{\bf{x}, \bf{y}}   \bf{\epsilon_y} \label{errx}\\
\frac{d\bf{\epsilon_y}}{dt} & = & \frac{\partial\bf{s}}{\partial \bf{y}}|_{\bf{y}}   \bf{\epsilon_y} \label{erry}
\end{eqnarray}
where $\bf{\epsilon_x}$ and $\bf{\epsilon_y}$ are the error vectors of variables $\bf{x}$ and $\bf{y}$, respectively. Their dynamics are determined by the properties of the entries of the Jacobian matrix, $\frac{\partial\bf{f}}{\partial \bf{x}}$,  $\frac{\partial\bf{f}}{\partial \bf{y}}$ and $\frac{\partial\bf{f}}{\partial \bf{x}}$ along the trajectory $\bf{x}, \bf{y}$. As the Eq. \ref{erry} does not depend on $\bf{x}$ and its error, the amplification is only determined by the instability properties of the tropical model, or in other words by its Lyapunov exponents, which have been computed using the standard Gram-Schmidt method which may be found in \cite{Kuptsov2011}. After a long integration on the system's attractor of 3000 years, we get the three exponents equal to $ \sigma_1= 0.000295$, $\sigma_2= -0.00000244$ and $\sigma_3= -0.00368$ day$^{-1}$. The solutions are chaotic and the typical time scale of error amplification is therefore equal to $\tau_1=1/\sigma_1=9.3$ years. This time scale is obviously well beyond the typical error amplification (and saturation) in the extratropical model, which is of the order of a few days \cite{Vannitsem2017}.

This situation implies that once the error is introduced initially in the tropical system, it stays almost the same for quite a while and can be viewed as a constant forcing in the Eq. \ref{errx}, which can then be viewed as a parametric error during the initial evolution of the error in the extratropics as
\begin{equation}
\frac{d\bf{\epsilon_x}}{dt}  =  \frac{\partial\bf{f}}{\partial \bf{x}}|_{\bf{x}, \bf{y}}   \bf{\epsilon_x} + \delta \bf{\mu}  \label{errxmu}
\end{equation}
where $\delta \bf{\mu} = \frac{\partial\bf{f}}{\partial \bf{y}}|_{\bf{x}, \bf{y}}   \bf{\epsilon_y} \approx$ constant. This feature implies that provided the initial condition error in the extratropics is going to zero while the initial error in the tropical module is different from zero, the mean square error will amplify quadratically initially as discussed in details in \cite{Vannitsem2002,Nicolis2003,Nicolis2007}.

This point will be taken up again in the next section while investigating the error dynamics in the coupled tropics-extratropics system.

\subsection{long time scales}
\label{longtime}

In \cite{Vannitsem2023}, it was shown that the mean square error evolution is quickly reaching large values in the extratropics which does not allow to use the linearized equations discussed in the previous section. The long term evolution of the error could still however show an amplification on long time scales which depends on the amplitude of the initial perturbations in the tropical module as illustrated in Fig. 4 of \cite{Vannitsem2023} for the first barotropic mode $\psi_{a,1}$ of the atmosphere. This slow increase of the mean square error is accompanied with a decrease of the correlation skill as illustrated in Fig. 5 in the same paper. Interestingly, when there is no error in the initial state of the tropical model, after a first quick decrease of the correlation, a plateau is reached reflecting the impact of the tropical forcing.

\begin{figure}
\centering
\includegraphics[width=140mm]{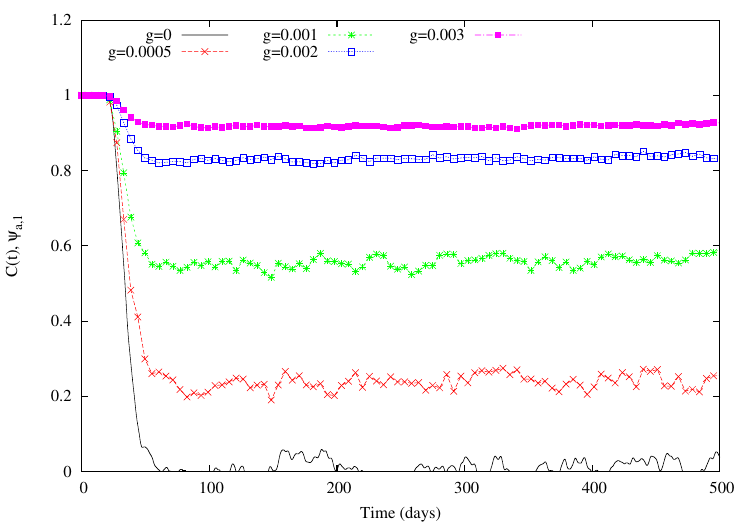}

\caption{Evolution of the correlation skill between the forecasts and the reference when non initial condition errors are present in the tropical forcing. Different values of the coupling parameter $g$ are used. The Figure is an adaptation of Figure 3, panel b of \cite{Vannitsem2023}.}

\label{perfectpredict}
\end{figure}

Figure \ref{perfectpredict} displays the correlation skill obtained for different values of the coupling parameter $g$ when no error in the initial state of the tropical module are present. These curves were qualified as "Perfect Potential Predictability" as it is maximum level of predictability within the nonautonomous extratropical atmosphere which could be reached due to the presence of the tropical module. Since errors are always present in the forcing of the extratropics, this level is never reached as illustrated in \cite{Vannitsem2023}.

\subsection{Experimental setup}
\label{setup}

When forecasting a system, it is important to provide information on the uncertainties. A traditional way to provide such information in weather forecasting and climate predictions and projections is to generate a set of trajectories originating from different initial conditions within the uncertainty range defined from observational uncertainties or within the data assimilation process. This operation leads to an ensemble forecast from which uncertainty ranges or probabilistic forecasts can be generated. In the current work, such an ensemble is also generated using 50 members. In the following, we will also use a set of 4000 different initial conditions along the reference trajectory of the system in order to get statistics independent of the initial state considered. The statistics that will be used to evaluate the quality of the forecasts at any time $t$ are

\begin{itemize}
    \item the Mean Square Error (MSE) of the forecast, $\psi_{a,1,for}$, which is given by
    \begin{equation}
        MSE=< (\psi_{a,1,for} - \psi_{a,1,ref})^2 >
    \end{equation}
    where $\psi_{a,1,ref}$ is the observed or reference value, and $\psi_{a,1,for}$, the forecast which could be either a single member forecast or the ensemble mean $\overline{\psi_{a,1}}$. The brackets, $<.>$, denotes an average over a large sample of different forecasts.
    \item the covariance between the forecast and the observed or reference value,
     \begin{equation}
        Cov(\psi_{a,1,for},\psi_{a,1,ref})=< (\psi_{a,1,for}-< \psi_{a,1,for}>) (\psi_{a,1,ref}-<\psi_{a,1,ref}>) >
    \end{equation}
    and the correlation skill
    \begin{equation}
        C(\psi_{a,1,for},\psi_{a,1,ref})=\frac{Cov(\psi_{a,1,for},\psi_{a,1,ref})}{ \sqrt{< (\psi_{a,1,for}-< \psi_{a,1,for}>)^2> <(\psi_{a,1,ref}-<\psi_{a,1,ref}>)^2>}}
    \end{equation}
    and as for the MSE, the forecast could be either a single member forecast or the ensemble mean
    \item Interestingly there is a nice relation between the correlation skill and the MSE of the forecasts as follows
    \begin{eqnarray}
        MSE & = & < (\psi_{a,1,for}-< \psi_{a,1,for}>)^2> + <(\psi_{a,1,ref}-<\psi_{a,1,ref}>)^2> \nonumber  \\ & & - 2 Cov(\psi_{a,1,for},\psi_{a,1,ref}) + (< \psi_{a,1,for}>-< \psi_{a,1,ref}>)^2
        \label{MSEdecomp}
    \end{eqnarray}
\end{itemize} 
where the two first terms are the variances of the forecasts and reference trajectories, the third term is the covariance between them and the last term is the square of the bias.
This relation allows for getting access to the amplitude of the bias between the forecast and the reference trajectory. This will be used later to clarify the effect of the initialization of the tropical model. 

Note also that in the case of an ensemble forecast, when $\psi_{a,1,ref}$ is a single member forecast of the ensemble which is compared to the ensemble mean, $\overline{\psi_{a,1}}$, it is usually referred as the potential skill while when $\psi_{a,1,ref}$ is the reference/observed trajectory, it is referred as the actual skill \cite{Kumar2014}. 

In \cite{Vannitsem2023}, the initialization of the model forecasts was done assuming that the error in observing both the tropics an the extratropics is a uniformly distributed random noise for all variables. It is also further assumed that we know this distribution exactly and one can therefore generate an appropriate ensemble forecasting system based on the same distribution as the one used to generate the uncertainty of the observations. The ensemble forecasts are therefore done as follows: random perturbations mimicking the uncertainty on the initial conditions in the extratropics is introduced from a uniform distribution of mean 0 and range $[-5. 10^{-7}: 5. 10^{-7}]$. This amplitude is very small typically of the order of 0.001 $\%$ of the variability of $\psi_{a,1}$. This mimics a situation for which the initial conditions are very well known in the extratropics allowing for having access to the different phases of the error dynamics.

The tropical dynamics is  affected by initial condition errors, that will in turn affect the forecasts in the extratropics. In order to mimic this feature, a random initial error is also introduced in the initial conditions
of the tropical model, generated from a uniform distribution of zero mean and a range $[-0.025:0.025]$, a typical case already addressed in \cite{Vannitsem2023}. Three different experiments will be done here: A first one in which the perturbations are introduced at random in any direction in phase space; a second experiment in which random perturbations are introduced selectively in the two-dimensional subspace spanned by the first and second Lyapunov vectors; and a third experiment in which the random perturbations only affect the first Lyapunov vectors. The second experiment would correspond to the projection of initial error in the unstable subspace of the system, a feature shared by many data assimilation system, \cite{Pires1996, Swanson1998, Carrassietal2022}. The third experiment would correspond to the case where the error explores only the unstable direction. Moreover, the amplitude of the perturbations along these different directions are adjusted in such a way to have a comparable MSE at initial time in the tropical forcing.

\section{Predictability in the extratropical module}
\label{sec:predictability}

\subsection{Error evolution}

\begin{figure}
\centering
\includegraphics[width=140mm]{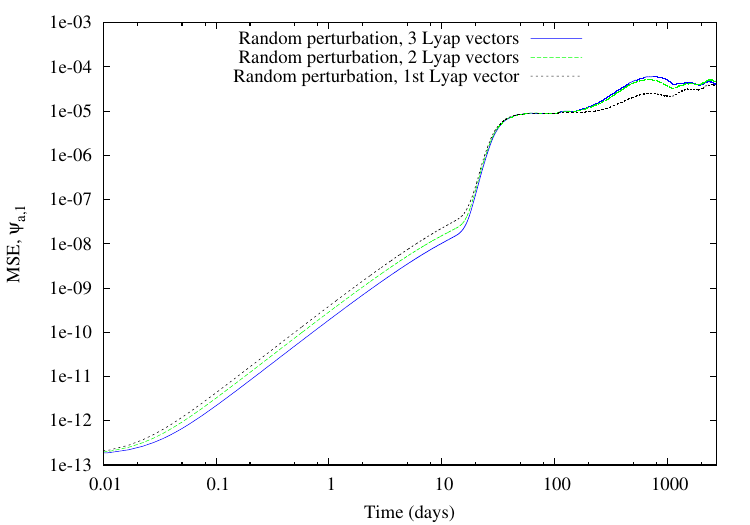}

\caption{MSE evolution for the variable $\psi_{a,1}$ . Three curves are displayed for random perturbations imposed along the three Lyapunov vectors for a initial error amplitude in the tropical model $PT=0.05$ (blue continuous line), along the two dominant Lyapunov vectors for a initial error amplitude in the tropical model $PT=0.05 \sqrt{3/2}$ (green dashed line) and along the most unstable Lyapunov direction for a initial error amplitude in the tropical model $PT=0.05 \sqrt{3}$ (black dotted line). The slight rescaling of the perturbations ensures a similar initial MSE for the three experiments.}

\label{errorevol}
\end{figure}

Let us now start with the description of the MSE evolution for the three first experiments described above in section \ref{setup}. Figure \ref{errorevol} displays the MSE between the variable $\psi_{a,1}$ of the model in a doubly logarithmic scale in order to clarify the initial error amplification. For the first experiment in which the error is introduced in all directions in phase space (blue continuous curve), the error slowly amplifies and then starts to follow a power law with an exponent close to 2. The feature is reminiscent of the error amplification of model parametric errors as discussed in section \ref{shorttime}. After 10 to 20 days, the error starts to increase following a more standard picture, with a exponential-like evolution followed by a linear growth and then reaching a saturation level. The saturation level, however, is far from simple as after a plateau persisting up to 200 days, the error starts again to increase and oscillates until the end of the experiment. The latter evolution at saturation has already been reported in \cite{Vannitsem2023} for purely random initial perturbations in the tropical model. 

Now let us compare with the results shown when perturbing in the unstable subspace defined by the two first Lyapunov vectors and along the most unstable Lyapunov vector. Differences are found for short times and for long times. For short times, a slight difference in the way the error reaches the quadratic amplification is experienced, related to the difference in convergence of the tropical solution toward its attractor. More interestingly is the error evolution after 200 days of forecast. In this case perturbing only along the dominant Lyapunov vector is reducing the error in the long term. This very non-intuitive behavior will be discussed in more details later, but is related to the low model bias induced by the tropical model in the third experiment.  

\subsection{Ensemble forecast long-term predictability}

Let us now turn to the analysis of ensemble forecasts. An important requirement in ensemble forecasts is reliability. This property consists in assuming that the reference trajectory (i.e. the observations) is indistinguishable to any member of the ensemble forecast. That means that any member could be exchangeable with the reference trajectory, and therefore, any score computed either with the reference trajectory or this new selected member should be the same, provided that we know exactly the properties of the initial conditions and that the model does not display any uncertainty. In a way a perfect model setting, even if uncertainties are present in the initial states. 

These considerations suggest therefore that the MSE against the ensemble mean of either the observation/reference trajectories or an ensemble member should be the same. In other words the spread of the ensemble should be equal to the root of the MSE between the observation and the ensemble mean (see e.g. \cite{leutbecher2008}). A similar consideration can be applied to the correlation skill: the correlation skill between the ensemble mean and the reference trajectory or the one between the ensemble mean and any member of the ensemble should be equivalent. In climate science, the correlation skill between any member and the ensemble mean is known as the potential predictability or skill \cite{Kumar2014}, and is considered as the reference to reach with any forecasting systems, provided the model is representing properly the system of interest. 

\begin{figure}
\centering
\includegraphics[width=140mm]{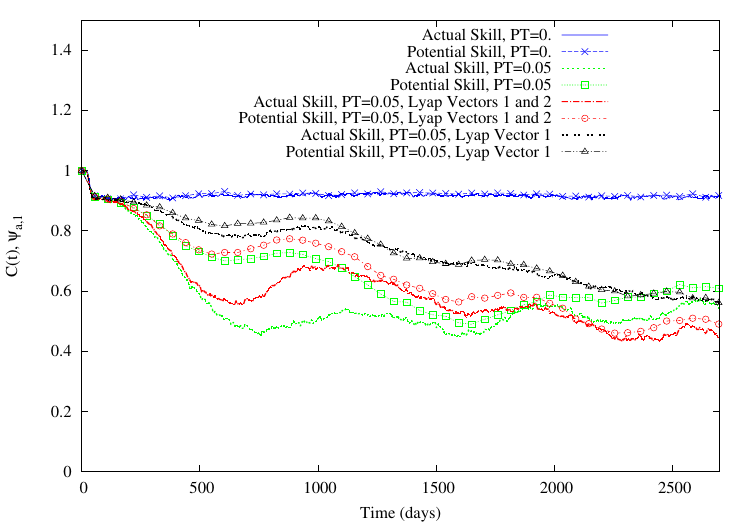}

\caption{Correlation skill for $\psi_{a,1}$. Two different skills are displayed: the actual skill comparing the reference trajectory with the ensemble mean as lines without symbols, and the potential skill comparing a random ensemble member with the ensemble mean as lines with symbols. The results for four experiments are displayed: (i) without error in the initial states of the tropical model, reflecting the perfect potential predictability of $\psi_{a,1}$ (blue curves and crosses); (ii) with random initial errors of amplitude $PT=0.05$ along the three Lyapunov vectors (green curves and open squares); (iii) with random initial errors of amplitude $PT=0.05 \sqrt{3/2}$ along the two dominant Lyapunov vectors (red curves and open circles); (iv) with random initial errors of amplitude $PT=0.05 \sqrt{3}$ along the first Lyapunov vector (black curves and open triangles).   }

\label{ensmeanspread}
\end{figure}

Figure \ref{ensmeanspread} shows the correlation skill of the ensemble mean as compared to the reference trajectory and a member of the ensemble in several cases. The blue curves show display these two quantities when no errors are present in the initial state. In such a case, both curves superimpose each other indicating that the actual skill is equal to the potential skill. The level reached after the quick initial decrease is the same as in Fig. \ref{perfectpredict} for $g=0.002$. When random perturbations are affecting all directions in phase space, the two curves (green dotted curve for the actual skill and the open square curve for the potential skill) are now well apart beyond 200 days. This feature was already shown in \cite{Vannitsem2023}, but without the proper explanation that will be provided below. When random perturbations are confined to the unstable subspace at initial time in the tropical model, the picture is slightly different with an overall improvement (red dashed-dotted curve for the actual skill and the open circle red curve for the potential skill), but the two curves are still quite distinct. Finally, the black curves (dotted for the actual skill and the open triangles for the potential skill) are obtained when perturbations are only affecting the dominant direction of instability, showing now a situation that we would expect with the actual skill very close to the potential skill in a perfect experimental setup framework (perfect model and perfect knowledge of the error distribution in the initial states). So it suggests that perturbing along the unstable direction (with the same amplitude of the MSE at initial time to the other experiments) allows for improving the skill of the forecasts. 

\begin{figure}
\centering
\includegraphics[width=140mm]{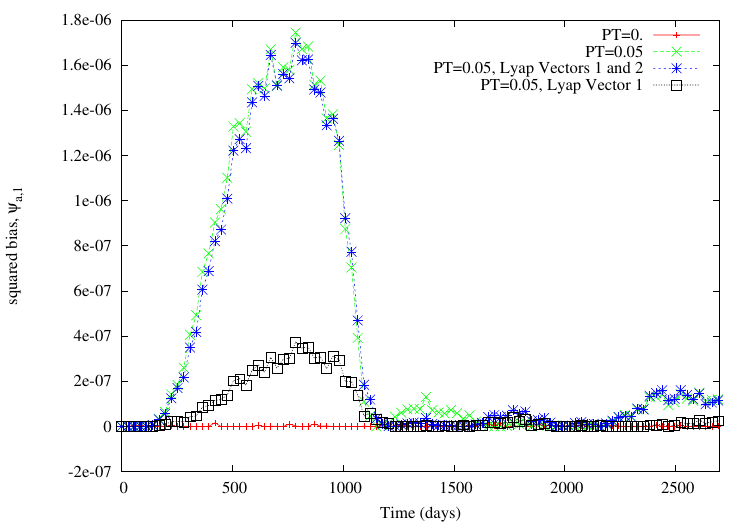}

\caption{Squared bias $(< \psi_{a,1,for}>-< \psi_{a,1,ref}>)^2$ for $\psi_{a,1,for}=\overline{\psi_{a,1}}$, the ensemble mean of an ensemble of 50 members. The results for four experiments are displayed: (i) without error in the intial states of the tropical model, reflecting the perfect potential predictability of $\psi_{a,1}$ (red curve and plusses); (ii) with random initial errors of amplitude $PT=0.05$ along the three Lyapunov vectors (green curve and crosses); (iii) with random initial errors of amplitude $PT=0.05 \sqrt{3/2}$ along the two dominant Lyapunov vectors (blue curve and stars); (iv) with random initial errors of amplitude $PT=0.05 \sqrt{3}$ along the first Lyapunov vector (black curve and open squares).}

\label{bias}
\end{figure}

How this highly counterintuitive result may happen? This question can now be understood by investigating first the MSE decomposition of Eq. \ref{MSEdecomp}, and isolating the squared bias between the ensemble mean, $\psi_{a,1,for}=\overline{\psi_{a,1}}$, and the reference trajectory, which is displayed in Fig. \ref{bias}. The blue curve with pluses without any initial condition error in the forecasting of the tropical system is very close to 0 (and almost invisible on the picture). The case where random perturbations in all directions are applied shown by the green dashed and pluses curve shows now a large squared bias whose evolution shows a damped oscillation. A very similar picture is found when perturbations are introduced in the unstable subspace defined by the two first backward Lyapunov vectors (short-dashed black curve with star symbols). The results are however completely different for perturbations introduced along the most unstable direction (dotted black curve with open squares). In this case, the squared bias is considerably reduced, explaining the improvement in the correlation skill and the MSE. The last curve (dashed-dotted green curve with full squares) shows the squared bias when perturbing along the tangent to the trajectory which does not show any bias. Note that the latter direction in phase space also correspond to the covariant Lyapunov vector associated with the 0 exponent of the model for a three-variable chaotic system. If the same analysis is done when the reference trajectory is replaced by a member of the ensemble, no substantial biases are found whatever the experiment. 

\begin{figure}
\centering
\includegraphics[width=140mm]{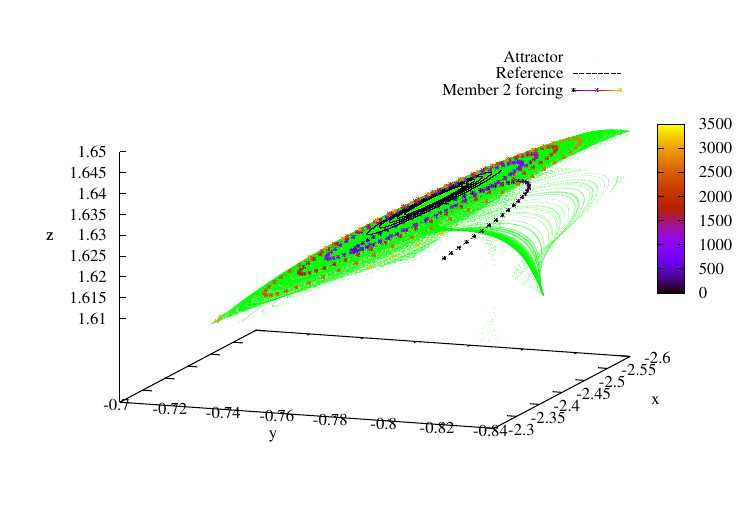}

\caption{A zoom on a portion of the attractor of the solutions of the tropical module of the system (green dots) and two trajectories of a specific ensemble forecast, namely the reference trajectory (black dashed curve) and a perturbed trajectory with random initial errors of amplitude $PT=0.05$ along the three Lyapunov vectors (colored curve with symbols). The color coding is related to the time evolution starting from black at $t=0$ days.}

\label{convergence}
\end{figure}

These results suggest that the key element inducing a reduction of actual skill and its divergence from the potential skill is the way the solutions converge toward the attractor of the system. In order to understand this, we plot a portion of the reference attractor of the tropical model (green dots) in Fig. \ref{convergence}, together with the reference trajectory at a specific initial time (dashed black line) lying on the reference attractor and the perturbed forecast as the colored trajectory with the star symbols. This perturbed trajectory converges progressively back to the attractor with a time scale of the order of the inverse of the amplitude of the third Lyapunov exponent, $271$ days. As illustrated on this picture, the perturbed solution reaches a trajectory on the attractor with a larger amplitude of oscillation than the reference trajectory. This modification of amplitude of oscillation is associated with a longer period of oscillation, and therefore an oscillating difference is generated. Such a behavior is also experienced for other trajectories. We therefore believe that this difference of amplitude and period of the forcing is the reason why a bias is induced in the forecasting of the extratropical evolution.   

\section{Conclusions}
\label{sec:conclusions}

The impact of initial condition error structure in the tropics on the long term predictability in the extratropics is investigated in the tropical-extratropical reduced-order coupled system built in \cite{Vannitsem2022, Vannitsem2023}. The key conclusions of the analysis are
\begin{itemize}
    \item Forecast biases could be induced in the extratropics even if the model is perfectly known and the nature of the initial uncertainties well controlled. Our conjecture is that this feature is related to the convergence back to the attractor of the tropical model, which induces a non-trivial response to the shift in the phase of the signal forcing in the extratropics. The pure random forcing along all directions is inducing a strong bias which considerably affects the forecasts on the long term
    \item A highly counterintuitive result is found in which perturbing along the unstable directions (with the same amplitude as in the other experiments) in the phase space of the tropical model improve the long term forecast of the forced extratropical component.
    \item The equality of the potential skill and the actual skill is only ensured provided these biases are considerably reduced
\end{itemize}

A first consequence of these findings is that data assimilation should ensure that the error in the tropical pacific is aligned in the unstable subspace of the tropical attractor, in such a way to avoid strong convergence toward the attractor that would affect the recharge-discharge oscillation of the tropical pacific. This can most probably be ensured by using an appropriate data assimilation system in the tropics such as the ones discussed in \cite{Carrassietal2022}. A second consequence is that forecast biases in forced systems related to the type of initial errors introduced in the forcing system should be carefully explored before assuming that biases are associated with the presence of uncertainties in the model structure. Several sources of biases could indeed be present altogether, either from the initial condition error structure as well as the model structure.

A similar type of analysis should be performed in a more realistic setting, either in an intermediate order model or in a state-of-the-art global circulation model, in which in particular the seasonality is affecting the occurrence of the ENSO events. In this context, different types of initial errors projecting along the stable and unstable manifolds may be introduced, and check for the long term predictability in the extratropics.

So far, only the coupling with the tropical Pacific has been explored. It is known that other regions in the tropics display a low-frequency variability which can affect the extratropics, such as the Madden-Julian oscillation \cite{Barsugli2002, Straus2023}. A similar idealized setting could be interesting to put in place using a simplified model of the MJO, and check for the presence of biases in the forecasts in the extratropics.

\bibliographystyle{unsrt}  
\bibliography{teleconnections}

\end{document}